\documentclass[pre,twocolumn,showpacs]{revtex4}
\usepackage{epsfig}
\usepackage{amsmath}
\usepackage{amssymb}

\begin{document}

\title{Anomalous wave as a result of the collision of two wave groups on sea surface}
\author{V. P. Ruban}
\email{ruban@itp.ac.ru}
\affiliation{Landau Institute for Theoretical Physics RAS, Moscow, Russia} 

\date{\today}

\begin{abstract}
The numerical simulation of the nonlinear dynamics of 
the sea surface has shown that the collision of two
groups of relatively low waves with close but noncollinear 
wave vectors (two or three waves in each group with
a steepness of about 0.2) can result in the appearance
of an individual anomalous wave whose height is noticeably 
larger than that in the linear theory. Since such collisions 
quite often occur on the ocean surface, this scenario 
of the formation of rogue waves is apparently most typical 
under natural conditions.
\end{abstract}

%47.35.Bb  Gravity waves
%92.10.Hm  Ocean waves and oscillations

\maketitle

Anomalous waves (rogue waves, freak waves) constitute one of the most 
interesting phenomena in the hydrodynamics of the sea surface and create a serious
danger to ships. For this reason, they are actively studied 
(see, e.g., reviews [1-3], special issues of journals
[4, 5], and numerous references therein). Among
moderate waves, a single very steep wave more than
twice as high as neighboring waves sometimes suddenly appears; 
its height from the trough to the crest
reaches 30 m at a characteristic length of an ocean
wave of 200-250 m. Such an extreme wave can damage even large ships. 
Such a large wave exists for several periods and, then, disappears without traces. 
Several possible mechanisms of this phenomenon have
already been proposed. Within the model of potential
motions of a liquid in the absence of large-scale inhomogeneous currents, 
two mechanisms -- linear dispersion and nonlinear self-focusing 
(modulation instability [6, 7]) -- are the most remarkable. 
Let waves be characterized by the following typical parameters: wavenumber $\tilde k$ , 
amplitude  $\tilde A$,  and the number of waves in a group $\tilde\nu$. 
The linear mechanism providing random
spatio-temporal focusing is important in wave fields with small
Benjamin-Feir indices, $I_{\rm BF}\propto \tilde\nu \tilde k \tilde A \lesssim 1$ 
(relatively wide spectrum, short-range spatial correlations,
almost complete absence of coherent structures),
whereas the nonlinear mechanism is decisive in long-range correlated 
fields with the presence of coherent structures, where $I_{\rm BF} \gtrsim 1$. It 
is noteworthy that nonlinearity acts more strongly 
in long-crested (locally quasi-two-dimensional) than in short-crested 
(significantly three-dimensional) random wave fields [8-12].

The nonlinear mechanism of rogue waves was
studied in numerous works (in addition to the works
cited above, see, e.g., [13-16] for planar flows, [17-21]
for three-dimensional flows, and references therein).
On the contrary, this work concerns the case of small $I_{\rm BF}$
values because it is more typical under natural conditions. However, 
although the index $I_{\rm BF}$ averaged
over all wave groups is small, groups for which  $\nu k A \sim 1$
can exist with a certain probability. They play the main
role in the formation of anomalous waves.

It is usually accepted that the so-called second-order theory neglecting 
four-wave interactions describes well the following situation in a field with a
wide spectrum. Owing to dispersion, several waves
with different lengths and directions can randomly
become ``in-phase'' at a given place and at a given
time. Their main harmonics are added according to
the linear superposition principle. Nonlinearity only
``tunes'' higher harmonics (see reviews [1-3] and references therein). 
However, recent numerical experiments [22, 23] show that nonlinearity 
is more significantly involved in the formation 
of a rogue wave even at small $I_{\rm BF}$ values. 
In particular, nonlinearity elongates its 
crest and changes the ``lifetime''. This work
continues the numerical investigation of the effect of
nonlinearity on the characteristics of anomalous
waves formed through spatio-temporal focusing. The main
question that will be answered is as follows. If the initial data are such that dispersion, 
according to the linear theory, should result not in a single high wave but
in the collision of two groups each of two or three
waves, can nonlinearity distort a linear interference
pattern in the collision time so that a single anomalous
wave is formed? A positive answer will be obtained.
This result is quite nontrivial. It is important simply
because the highest waves in linear fields with a wide
spectrum (in particular, in the so-called crossing states
when two spectral maxima exist) appear through random collisions of more 
moderate wave groups, which always appear in the considered region, 
grow owing to focusing, and then disappear. Events where one high
group is directly focused are much rarer. For this reason, it is reasonable to study in detail 
the pair collision of wave packets on the water surface and the dependence of the properties 
of appearing anomalous waves
on the parameters characterizing the packets and their
mutual location at a conditional initial time. This is
the aim of the numerical experiments reported below.

In order to better understand the process of collision, 
it is useful to consider a simple variational model
that approximately describes the dynamics of an individual 
wave packet on deep water. Let $x$ and $y$ be the
horizontal coordinates, $z$ be the vertical coordinate, $g$
be the gravitational acceleration, ${\bf k}_0$ be wave vector
directed along the $x$ axis, $k_0=|{\bf k}_0|$ be the wavenumber,
$\omega_0=2\pi/T_0=\sqrt{gk_0}$ be the frequency of the carrier
wave, $v_{\rm gr}=(1/2)\sqrt{g/k_0}$ be the group velocity, and $A(x,y,t)$ 
be the complex envelope of the main harmonic. The vertical deviation of the free surface is
determined by the formula $z\approx\mbox{Re}[A\exp(i{{\bf k}_0}\cdot{\bf r}-i\omega_0 t)]$.
We begin with the corresponding nonlinear Schroedinger equation
\begin{equation}
2i\psi_{\hat t}+\psi_{\hat x \hat x}-\psi_{\hat y \hat y}+|\psi|^2\psi=0,
\label{NLS}
\end{equation}
written in the dimensionless variables  $\psi=k_0 A^*$,
$\hat t=\omega_0 t$, $\hat x=2k_0 (x-v_{\rm gr}t-x_0)$, and $\hat y=\sqrt{2}k_0 (y-y_0)$. 
Substituting the simplest Gaussian ansatz (see, e.g., [23-26] and references therein)
\begin{equation}
\psi=\sqrt{\frac{4N}{XY}}\exp\Big[-\frac{\hat x^2}{2X^2}-\frac{\hat y^2}{2Y^2} 
+i\frac{U\hat x^2}{2X} -i\frac{V\hat y^2}{2Y} +i\phi\Big]
\label{Gaussian}
\end{equation}
into the Lagrangian of the nonlinear Schroedinger equation
\begin{equation}
{\cal L}=\int(i\psi_{\hat t}\psi^*-i\psi\psi^*_{\hat t}
-|\psi_{\hat x}|^2+|\psi_{\hat y}|^2+|\psi|^4/2) d\hat x d\hat y
\label{L_NLSE}
\end{equation}
and performing the standard procedure of the derivation of variational equations, 
we obtain the homogeneous system
\begin{equation}
\ddot X=\frac{1}{X^3}-\frac{N}{X^2Y}, \qquad \ddot Y=\frac{1}{Y^3}+\frac{N}{Y^2X}
\label{XY_dyn}
\end{equation}
for the longitudinal, $X(t)$, and transverse, $Y(t)$, dimensions of the packet.
In this case,  $4\pi N=\int |\psi|^2 d\hat x d\hat y=$ const, 
$U=\dot X$, $V=\dot Y$.  It is worth noting that the
parameter $N$ at $Y\sim X$  is proportional to the square of
the local Benjamin-Feir index. If $N\sim 1$, dispersion
and nonlinear contributions on the right-hand sides
are of the same order of magnitude, so that the evolution of a wave 
packet cannot be divided into the linear
and nonlinear stages. The solutions of the system of
differential equations (4) are studied in detail [23, 24, 26].
 They describe three main stages of the evolution
of the wave packet. The stage of focusing corresponds
to the ballistic regime with $U\approx$ const $<0$ and $V\approx$ const $<0$. 
Then, the stage of the maximum compression of the packet occurs; the details and duration of
this stage depend on the initial conditions and can be
quite diverse at different $N$ values. Finally, the defocusing stage in the ballistic 
regime with $U\approx$ const $>0$ and  $V\approx$ const $>0$ occurs. 
It is worth noting that the applicability of the Gaussian variation ansatz to waves
on water even in its more general, off-diagonal variant is confirmed in [23].

\begin{figure}
\begin{center}
   \epsfig{file=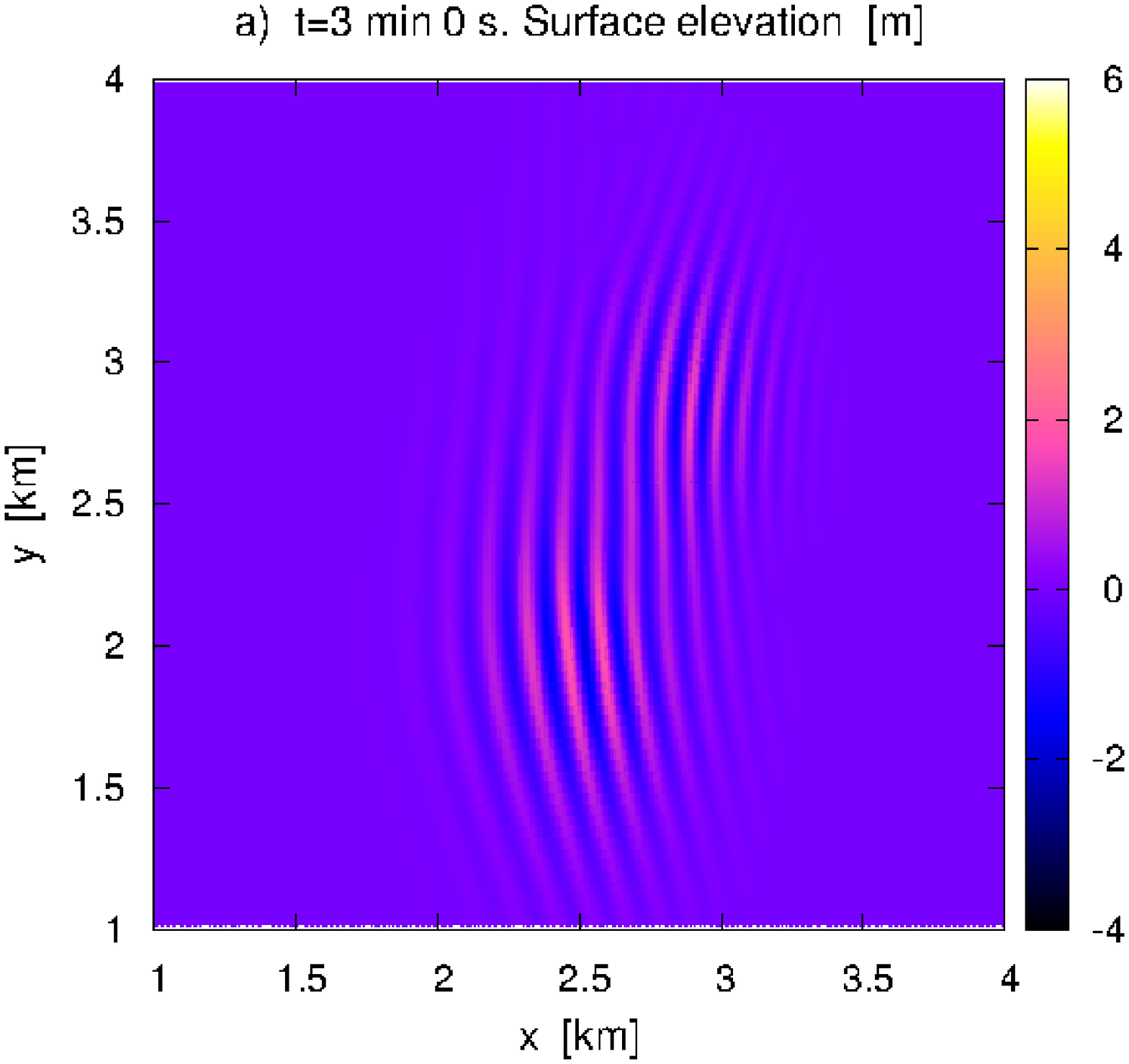, width=72mm}\\
\vspace{3mm}
   \epsfig{file=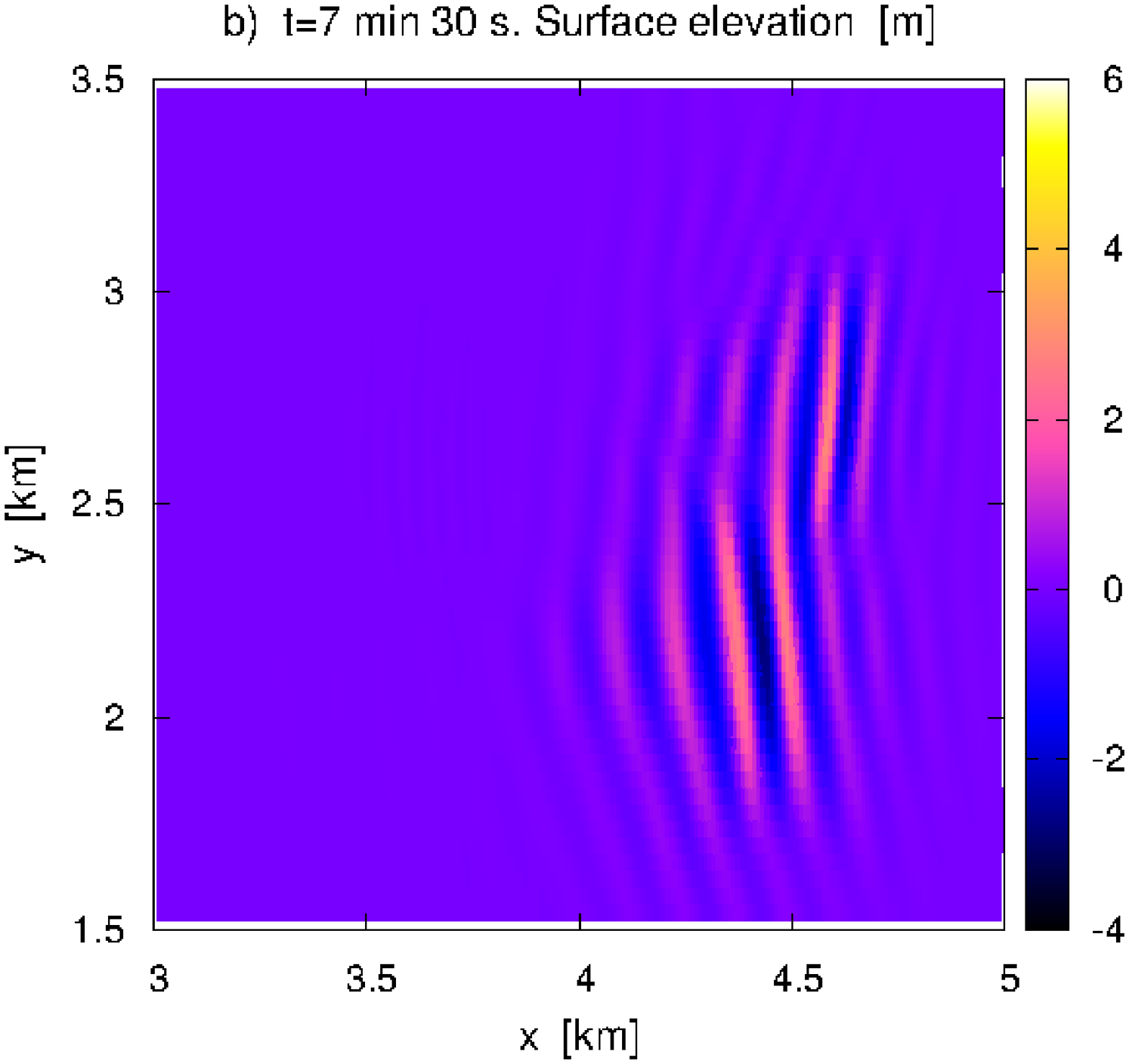, width=72mm}\\
\vspace{3mm}
   \epsfig{file=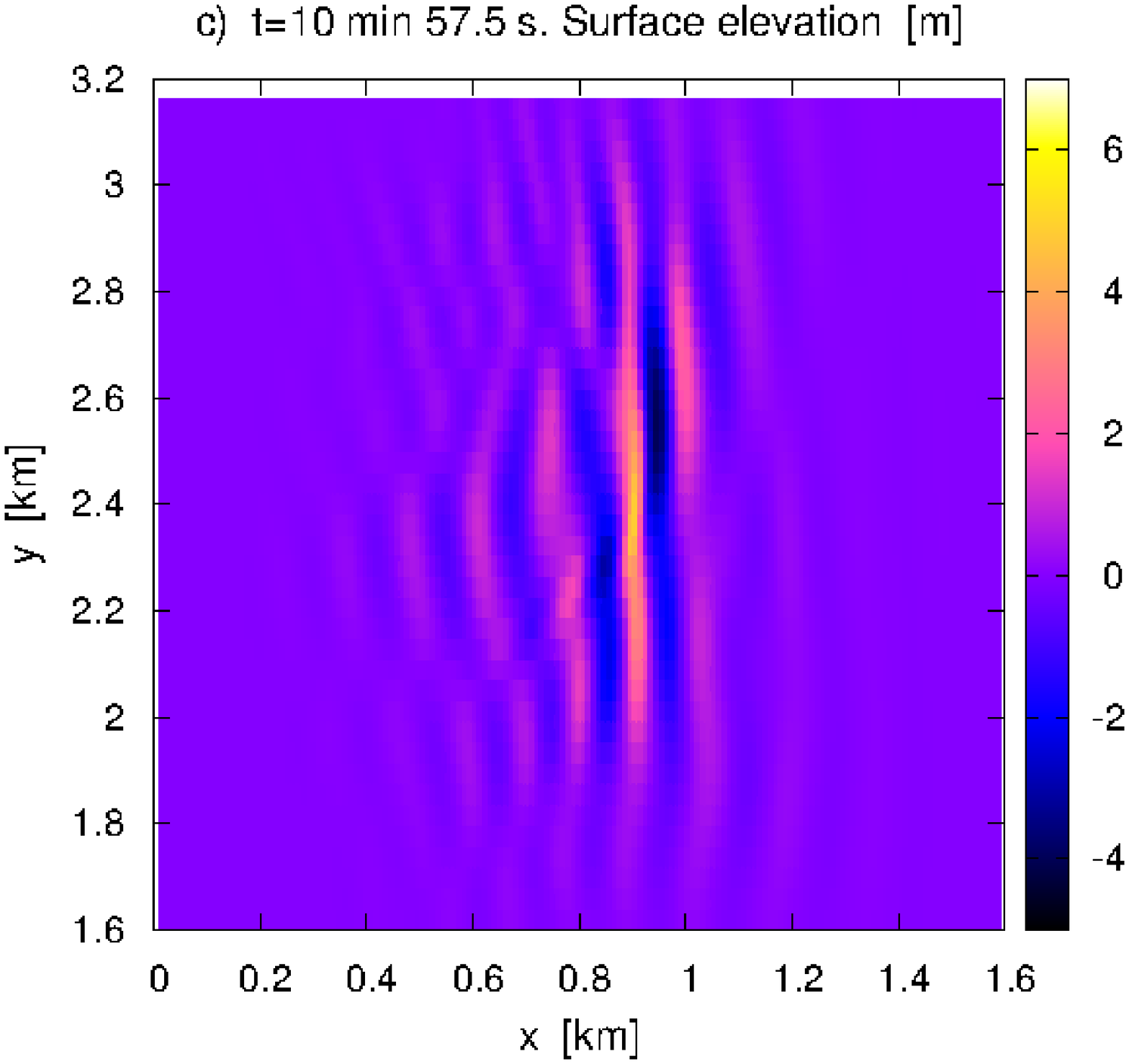, width=72mm}
\end{center}
\caption{Successive stages of the collision of
two wave packets: (a) focusing, (b) beginning of collision,
and (c) appearance of an anomalous wave. The parameters
are
$N=1.5$, $X_0=30$, $Y_0=40$, $U_0=-0.05$, $V_0=-0.10$, ${\bf k}_1=(40,4)$,  ${\bf k}_2=(50,-4)$, 
$\phi_1=\phi_2=0$, ${\bf r}_1=(0.5\pi,0.75\pi)$, ${\bf r}_2=(0.7\pi,1.18\pi)$.}
\label{STF76-maps} 
\end{figure}

\begin{figure}
\begin{center}
   \epsfig{file=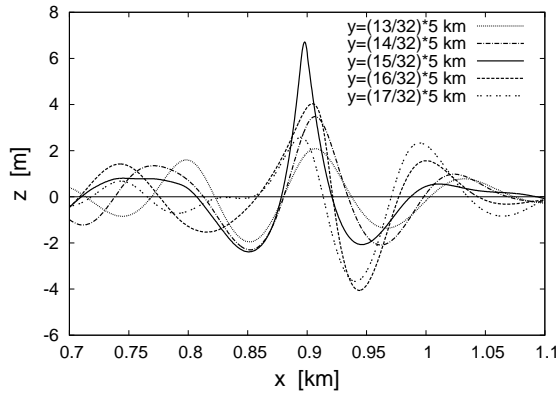,width=76mm}
\end{center}
\caption{Several wave profiles from Fig. 1c that confirm the
presence of a single anomalous wave. Similar profiles were
observed in many numerical experiments with random
wave fields (e.g., in [22, 28]), which indicates that the scenario 
of the formation of rogue waves under consideration
is realistic.
} 
\label{STF76-bw} 
\end{figure}

\begin{figure}
\begin{center}
   \epsfig{file=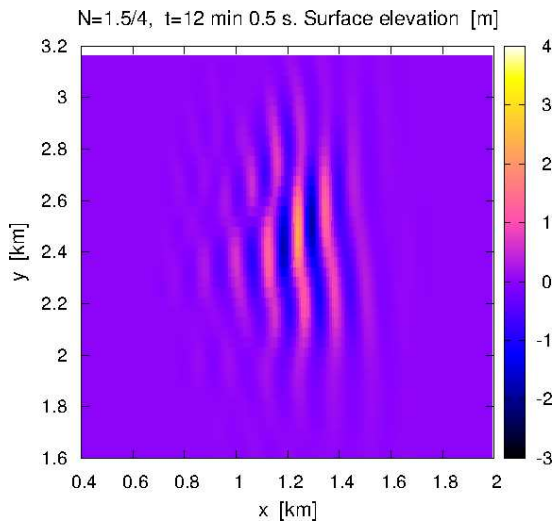,width=72mm}
\end{center}
\caption{Wave pattern at the collision of
packets with half the initial amplitude of that in Fig. 1. 
Significant differences from Fig. 1c are caused by an almost
linear character of the interaction in this case.
} 
\label{STF76A} 
\end{figure}

\begin{figure}
\begin{center}
   \epsfig{file=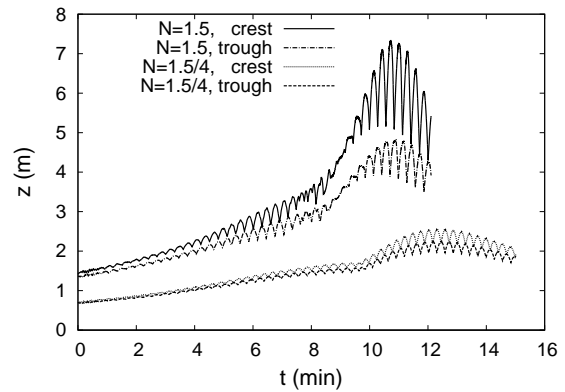,width=76mm}
\end{center}
\caption{Time dependence of the height of the highest crest
and the depth of the deepest trough at the same parameters
as in Fig. 1, as well as similar characteristics at half the initial amplitude. 
It is seen that the strong nonlinearity led
not only to the sharp ``top-bottom'' asymmetry owing to
higher harmonics but also to an about 2.5-fold increase in
the average amplitude of the maximum wave (half-sum of
envelopes) as compared to the weakly nonlinear regime.
} 
\label{STF76-kA} 
\end{figure}

In our numerical experiments, two nearly Gaussian packets oriented along the corresponding wave
vectors ${\bf k}_1$ and ${\bf k}_2$ (quite close, but noncollinear) at the
initial time were centered at the points ${\bf r}_1$ and ${\bf r}_2$. For
simplicity, the parameter $N$ and the initial values  $X_0$, $Y_0$, $U_0$, and $V_0$
were taken to be identical for both packets and corresponded to the ballistic focusing stage.
The initial positions of the centers ${\bf r}_1$ and ${\bf r}_2$ were chosen taking into 
account the group velocities so that collision occurs approximately at the stage of 
maximum compression. In this case, the initial dimensions
of the packets should be, on one hand, not too large so
that the packets do not significantly overlap and, on
the other hand, not too small so that dispersion has no
time to defocus them. Despite these constraints, the
parametric region of the initial conditions ${\bf r}_1-{\bf r}_2$ favorable 
for the formation of a large wave, which includes states with two maxima, is fairly wide.

The parameter $N$  was chosen such that the steepness of waves at the stage of maximum compression
was about 0.2 for each packet. In this case, an extremely nonlinear anomalous wave with a sharp
crest was formed at collision. With an increase in $N$, the wave developed higher, 
but computations were terminated because of the beginning of the crest breaking, 
which required overly small spatial and temporal resolutions in our numerical method. 
At smaller $N$  values, nonlinearity was insufficient.

To simulate the nonlinear dynamics of the free surface, the model of fullly nonlinear, 
weakly three-dimensional water waves described in [27, 28]
was used. The computational region was a square with a side of $2\pi$, 
with the periodic boundary conditions imposed. For better presentation, 
the results were rescaled to a square with a side of 5 km; as a result, e.g.,
a dimensionless wavenumber of 50 corresponds to a
wavelength of 100 m typical of the World Ocean.

\begin{figure}
\begin{center}
   \epsfig{file=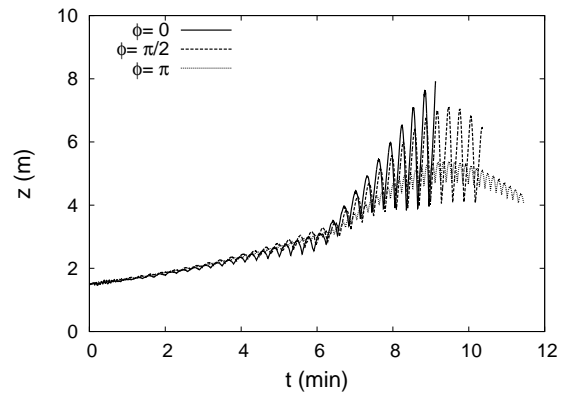,width=76mm}
\end{center}
\caption{Time dependence of the height of the highest crest
for three $\phi_1$ values at $\phi_2=0$ and the parameters
 $N=1.2$, $X_0=Y_0=30$,  $U_0=V_0=-0.07$, ${\bf k}_1=(40,6)$,  ${\bf k}_2=(40,-6)$, 
${\bf r}_1=(\pi,0.72\pi)$, ${\bf r}_2=(\pi,1.28\pi)$.
} 
\label{STF_02-crests} 
\end{figure}
\begin{figure}
\begin{center}
   \epsfig{file=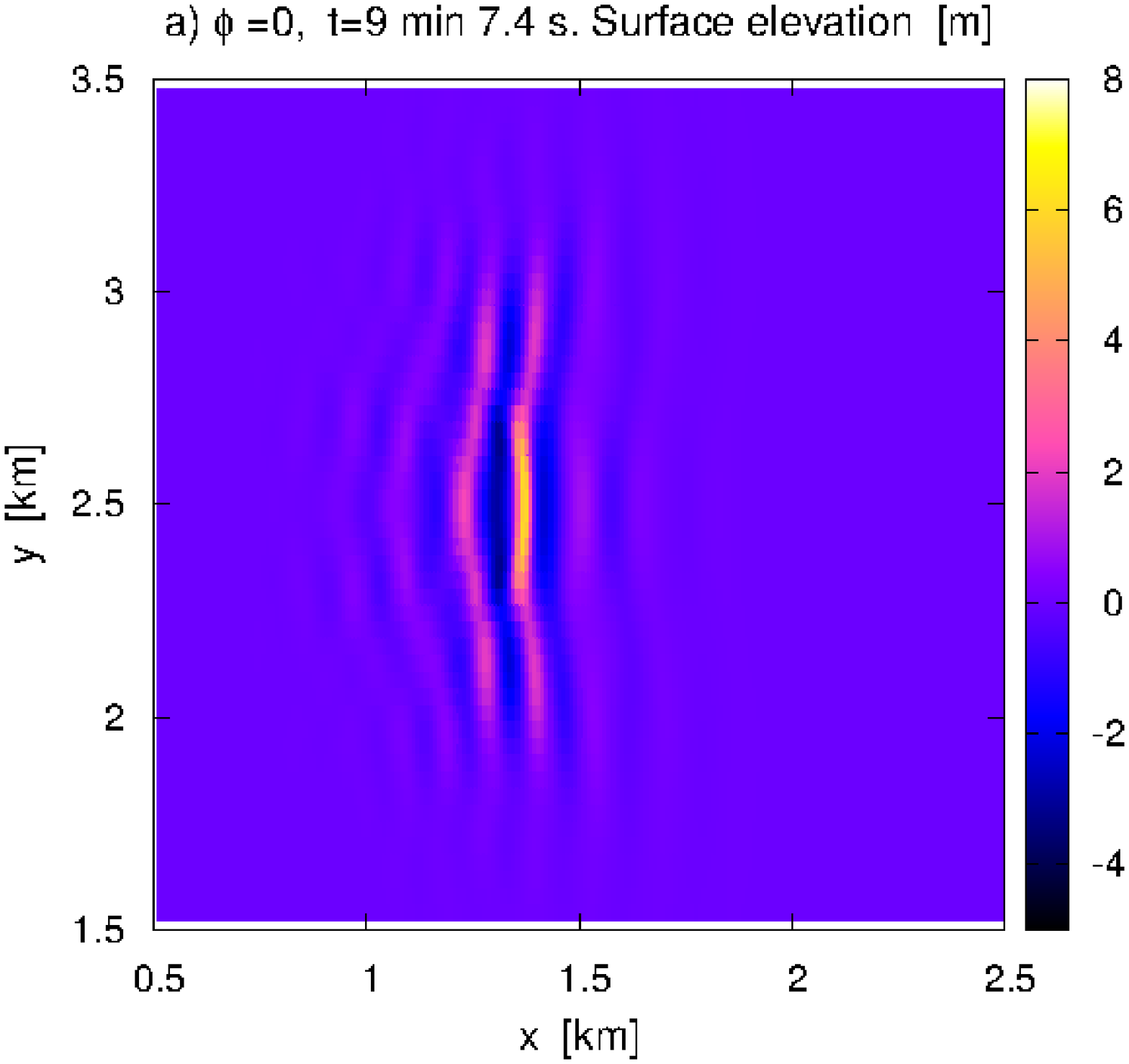,width=72mm}\\
\vspace{3mm}
   \epsfig{file=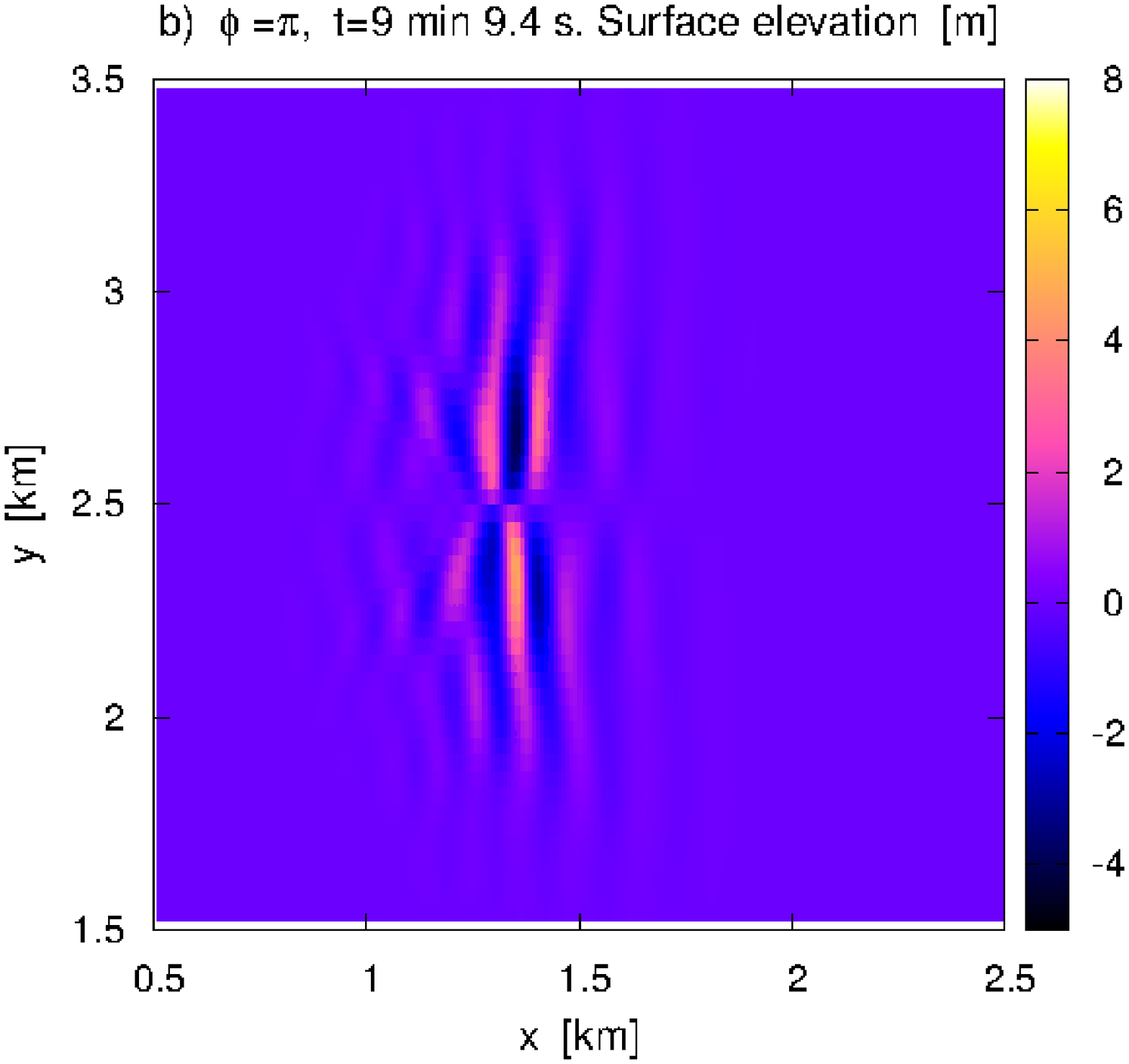,width=72mm}
\end{center}
\caption{Example of the effect of the phase
difference on the wave pattern at the collision of two wave
packets with the same parameters as in Fig. 5.} 
\label{STF_02-maps} 
\end{figure}

Figure 1 exemplifies the formation of an anomalous wave through the collision 
of two packets. At the first stage, when the packets are separated, their
almost independent focusing occurs. Then, the two
increased-amplitude regions begin to overlap (beginning of the collision). 
At this moment, each group consists
of two or three waves with a steepness of about 0.2. 
The length of crests is about three or four wavelengths and
the angle between their directions is about 0.2-0.3 rad. 
Further, the most interesting stage follows,
when nonlinearity is sharply enhanced and begins to
transform the interference pattern from two superimposed 
wave packets. Instead of a group of two or three
waves with the summarized amplitude, which would
observed in the case of the linear superposition of the
main harmonics, an extremely short group consisting
of nearly one wavelength, i.e., an individual anomalous wave, 
is formed, as is clearly seen in Figs. 1c and
2 (cf. Fig. 3, where the wave pattern at the collision of
packets with half the initial amplitude is shown). It is
important that the amplitude of the rogue wave is
noticeably larger than the sum of the amplitudes of
two groups even if only the main harmonic is taken
into account (see Fig. 4). The breather (oscillating)
type of this anomalous wave is remarkable: the state
with a high crest changes to the state with the deep
trough and vice versa, which also follows from Fig. 4.
The indicated property is due to the extreme shortness
of the group at the double difference between the
phase and group velocities. The period of these oscillations 
is approximately the doubled period of the
wave. Undergoing about ten of such oscillations, the
large wave expands in the transverse direction (i.e., its
crest is elongated), its amplitude decreases, and the
wave transfers to the final focusing regime (not shown in the figure).

Calculations with other initial conditions were also
performed. In particular, the strong effect of the phase
difference $\phi=(\phi_2-\phi_1)$ on the process of formation of
the anomalous wave (under identical other parameters) was revealed. 
If collision began so that the interference maximum of the linear theory had to pass
through the middle of the joined group, the anomalous wave was developed
more rapidly and was higher.
If the phase difference led to the interference minimum 
in the middle of the group, the wave was not so
high. Such a dependence of the behavior of the solution 
on the phase difference is exemplified in Figs. \ref{STF_02-crests} and \ref{STF_02-maps}.

Collisions of groups at $|{\bf k}_1|=|{\bf k}_2|$ in our numerical
experiments were more efficient than collisions with
noticeably different $|{\bf k}_1|$ and $|{\bf k}_2|$. This indicates that
transverse focusing is in some sense more important
than longitudinal focusing. In other words, three-dimensionality of the space 
is fundamentally important for the formation of rogue waves in random fields
with low Benjamin-Feir indices.

To summarize, a specific significantly nonlinear
mechanism of the formation of anomalous waves at
the collision of two quite small and not overly high
wave groups at an angle of about 0.2-0.3 rad has been
demonstrated. Unfortunately, our model cannot be
used to study collisions at larger angles.

\end{document}